\documentclass{article}
\usepackage{spconf}
\usepackage{cite}
\usepackage{amsmath,amssymb,amsfonts}
\usepackage{algorithmic}
\usepackage{graphicx}
\usepackage{textcomp}
\usepackage{xcolor}
\usepackage{multirow}

\begin{document}

\title{RelUNet: Relative Channel Fusion U-Net for Multichannel \\ Speech Enhancement}

\name{Ibrahim Aldarmaki$^1$, Thamar Solorio$^1$, Bhiksha Raj$^{1,2}$, Hanan Aldarmaki$^1$}
\address{Mohamed bin Zayed University of Artificial Intelligence$^1$ \\
Carnegie Mellon University $^2$
}

\maketitle

\begin{abstract}

Neural multi-channel speech enhancement models, in particular those based on the U-Net architecture, demonstrate promising performance and generalization potential. These models typically encode input channels independently, 
%and integrate the channels only prior to the final layer. 
and integrate the channels during later stages of the network.
In this paper, we propose a novel modification of these models %leveraging an end-to-end encoder-decoder mask-predicting architecture empowered 
by incorporating relative information from the outset, where each channel is processed in conjunction with a reference channel through stacking. 
% The integration of the GAT facilitates the modeling of different microphone channels within the array as graph nodes and uses an attention mechanism embedded within the GAT to adaptively fuse information from different channels, thereby capturing crucial spatial information and enhancing the overall speech signal quality. 
This input strategy %better models microphone channels within the array exploiting
exploits comparative differences to adaptively fuse information between channels, thereby capturing crucial spatial information and enhancing the overall performance.
The experiments conducted on the CHiME-3 dataset demonstrate  improvements in speech enhancement metrics across various architectures. 
%clarity, noise reduction, and generalization compared to traditional methods.
\end{abstract}


\section{Introduction}
Speech enhancement is a critical task in various applications, such as human-machine interfaces, mobile communications, and hearing aids \cite{das2021fundamentals, liu2014}. The objective is to improve the intelligibility and quality of speech that has been degraded by background noise and reverberation. This process becomes more complex in environments with multiple sound sources and reflections, necessitating advanced methods to isolate the target speech. The availability of multiple microphones (channels) help more accurately isolate sound sources that originate from different places.  Traditional multi-channel speech enhancement methods often rely on array processing techniques like spatial filtering, also known as beamforming. These methods utilize spatial information, such as the angular position of the target and the configuration of the microphone array, to distinguish the desired speech from noise \cite{van1988}. While beamforming approaches such as the minimum variance distortion-less response (MVDR) beamformer can effectively suppress unwanted noise, they depend heavily on prior accurate estimation of spatial information. This reliance can limit their performance in noisy and reverberant conditions where precise spatial information is difficult to obtain \cite{capon1969}.
In recent years, deep learning techniques have shown promise in leveraging data to discern patterns in multi-channel speech input. %addressing the limitations of traditional methods. 
Convolutional neural networks (CNNs) and U-Nets, in particular, have been effectively used for speech enhancement tasks due to their ability to learn complex local and global features from data. Originally introduced for image segmentation, U-Nets have demonstrated positive results in processing time-frequency segments of multichannel speech signals using parallel processing with shared weights to handle different channels \cite{ronneberger2015, tan2018convolutional}. 
Additional layers and network components have been proposed to improve performance 
\cite{tolooshams2020channel}. 
Previously proposed neural speech enhancement models treat input channels independently, and merge them through later mechanisms. For example, in \cite{tzirakis2021multi}, the authors integrated a Graph Neural Network (GNN) at the bottleneck of a U-Net architecture to enable cross-channel information sharing. Similarly, in \cite{chang2021end}, cross-channel attention layers were used in  a transformer architecture to encode information between channels.

In this work, we describe \textbf{Rel}ative Channel Fusion \textbf{UNet}, RelUNet, a variation on the standard U-Net architecture for multi-chanel speech enhancement. 
Unlike previous neural architectures that process each channel independently until the last layers, RelUNet
incorporates relative information by stacking each channel with a reference channel, 
thus modelling relationships between the multi-channel signals early in the network. This modification allows both parallel data processing for optimized computing and parameter usage as well as cross-channel fusion for accurate enhancement. 
We tested the approach with the vanilla U-Net architecture as well as variations by inserting Graph Convolutional Networks (GCN) or Graph Attention Networks (GAT) between the encoder and decoder, as done in  \cite{chau2024novel}.
We used the CHiME-3 dataset \cite{barker2015chime3} to demonstrate the effectiveness in both simulated and real-world noisy environments.

\section{Preliminaries}

\subsection{Problem Formulation}
In multichannel speech enhancement, we estimate a clean speech signal from noisy recordings captured by multiple microphones. Let $\textbf{x}_m \in \mathbb{R}^N$, where $N$ represents the number of samples in the signal, be the signal recorded by microphone $m$ at sample $n = 1, 2, ..., N$, which can be expressed mathematically as: 
\begin{equation}
    \textbf{x}_m[n] = \textbf{g}_m[n]\textbf{s}[n-n_m] + \textbf{n}_m[n]
\end{equation}

where $\textbf{g}_m$, $\textbf{s}$, $n_m$, and $\textbf{n}_m$ represent the gain, clean speech signal, time delay,and captured noise, respectively,
 with $m = {1,2,...,M}$, where $M$ is the number of microphones. Such time-domain signals can be transformed into the frequency domain using the Short-Time Fourier Transform (STFT) , which is given by:

\begin{equation}
    \textbf{X}_m(f,t) = \textbf{h}_m(f,t)\textbf{S}_m(f,t) + \textbf{N}_m(f,t)
\end{equation}
where $\textbf{X}_m$, $\textbf{S}_m$, and $\textbf{N}_m$ are the STFT of $\textbf{x}_m$,  $\textbf{s}_m$, and $\textbf{n}_m$,  respectively,  $\textbf{h}(f,t) = [\textbf{G}_1(f,t)  e^{(-j2\pi f f_s \tau _1 /K)}, ...,\\ \textbf{G}_M(f,t)  e^{(-j2\pi f f_s \tau _M /K)}]$ represents the steering vector, where $\textbf{G}_m$ is the STFT of $\textbf{g}_m$, $f$ is the frequency bin, $f_s$ is the sampling frequency, and $t$ is the time segment.

\subsection{Traditional Beamforming}

The MVDR beamformer is  a widely used adaptive beamforming technique; its weights are obtained by solving a constrained optimization problem that ensures that the beamformer maintains a distortionless response in the direction of the desired signal. The solution to this optimization problem is given by the following closed-form expression: 
\begin{equation}
    \textbf{w}_{MVDR}(f) = \frac{\textbf{R}_{nn}^{-1}(f,k) \textbf{h}(f,k)}
    {\textbf{h}^H(f,k)  \textbf{R}_{nn}^{-1}(f,k)  \textbf{h}(f,k)}
\end{equation}
where $\textbf{R}_{nn} \in \mathbb{R}^ {M \times M}$ represents the noise covaraince matrix.

Effective implementation of the MVDR beamformer requires accurate estimation of gain, Time Difference of Arrival (TDOA) and noise covariance matrix. Generalized Cross Correlation with Phase Transformation (GCC-PHAT), 
as defined in equation \ref{eqn:eqn_gcc_phat}, is a widely used method for robust TDOA estimation as $\hat{TDOA} = \underset{\tau}{argmax} (GCC-PHAT(\tau))$.

\begin{equation} \label{eqn:eqn_gcc_phat}
    GCC-PHAT(\tau) = \int_{-\infty}^{\infty}\frac{X_i(f)X^*_r(f)}{|X_i(f)X^*_r(f)|}  {e^{j2\pi f \tau}} df
\end{equation}

\subsection{Motivation}

The motivation behind our approach stems from a gap in how traditional signal processing techniques and modern deep learning architectures handle multichannel data. Signal processing methods typically incorporate the multichannel aspect early in the data processing pipeline, relying on techniques like cross-correlation, defined in equations \ref{eqn:eqn_gcc_phat} and \ref{eqn:cross_correlation_x1_x2}, or cross-spectral density, defined in equation \ref{eqn:eq_csd_raw_x1_x2} (which are related via the Wiener-Khintchin theorem \cite{wiener1930generalized, khintchine1934korrelationstheorie}, as shown in equation \ref{eqn:eq_wiener_khintchin_x1_x2}), 
to achieve accurate speech enhancement. However, most deep learning architectures previously explored do not facilitate this early-stage cross-channel fusion, and process each channel independently.
The proposed model incorporates such cross-channel information fusion, parallel data processing, and efficient parameter usage.

\begin{equation} \label{eqn:cross_correlation_x1_x2}
R_{x_1,x_2}(\tau) = \int_{-\infty}^{\infty} x_1(t) x_2(t + \tau)^* \, dt
\end{equation}

\begin{equation} \label{eqn:eq_csd_raw_x1_x2}
S_{x_1,x_2}(f) = X_1(f) X_2(f) ^*
\end{equation}

% Wiener-Khintching Theorem
\begin{equation} \label{eqn:eq_wiener_khintchin_x1_x2}
S_{x_1,x_2}(f) = \int_{-\infty}^{\infty} R_{x_1,x_2}(\tau) e^{-j 2 \pi f \tau} \, d\tau
\end{equation}

\section{Proposed Method}
\begin{figure*}[h]
\centering
\includegraphics[width=0.6\textwidth]{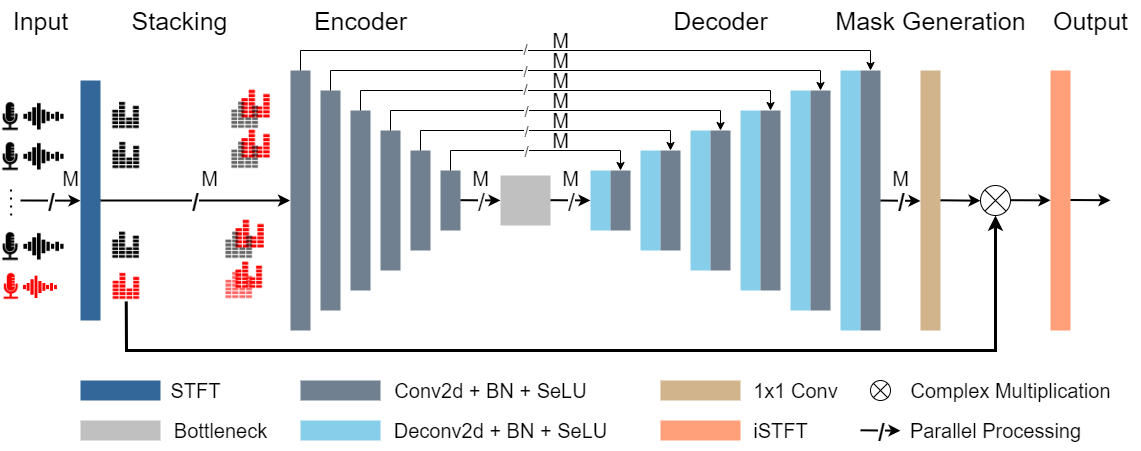}
\caption{Illustration of the proposed multi-channel speech enhancement method using a RelUNet.}
\label{fig:proposed_method}
\end{figure*}

\subsection{Input Representation}

In RelUNet, we first transform time-domain speech signals, $\textbf{x}_i \in \mathbb{R}^N$ where $N$ represents the number of samples, into time-frequency representations using the STFT. 
Then, we stack the real and imaginary parts of the STFT sequence of an input channel signal, $\textbf{X}_i \in \mathbb{C}^{F \times T}$, where $F$ is the number of frequency bins and $T$ is the number of time frames, with the corresponding STFT components from a reference signal, $\textbf{X}_r \in \mathbb{C}^{F \times T}$, as shown in Figure \ref{fig:proposed_method}. The resulting channel input, $\textbf{Z}_i \in \mathbb{R}^{4 \times F \times T}$, is expressed as:

\begin{equation}
    \textbf{Z}_i = \begin{bmatrix}
        Re(\textbf{X}_i),
        Im(\textbf{X}_i),
        Re(\textbf{X}_r),
        Im(\textbf{X}_r) 
    \end{bmatrix}
\end{equation}
where $Re(.)$ and $Im(.)$ denote the real and imaginary parts of the STFT, respectively. Considering all M input channels results 
in input $\textbf{Z} = \{\textbf{Z}_i\}_{i=1}^{M} \in \mathbb{R}^{M \times 4 \times F \times T}$.

This modification enables the integration of relative information in the microphone array, anchored by a reference channel, throughout the network, enhancing the representational capacity of the subsequent computations.

\subsection{Feature Representation and Mask Generation}
The U-Net architecture was used to extract feature representations from the complex STFT inputs. The encoder-decoder structure with skip connections helps preserving spatiotemporal information throughout the network. The encoder maps the stacked input channels into latent representation $\boldsymbol\zeta  \in \mathbb{R}^{M \times d \times \Tilde{F} \times \Tilde{T}}$ , where d is the number of filters/kernels used
in the encoder’s last convolutional layer and $\Tilde{(.)}$ represents the reduced dimensions due to encoder's down-sampling.

The decoder then upsamples these feature maps, constructing the input to the mask generating network. The output of the decoder is $\textbf{D}_{out} \in \mathbb{R} ^ {M \times d \times F \times T}$. 
The input to the mask generating network is a concatenation of $\textbf{D}_{out}$ across all $M$ channels in the second dimension, which can be represented as $Concat({\{\textbf{D}_i\}}^M_{i=1}) \in \mathbb{R}^{MK \times F \times T}$. The mask generating network uses a 1x1 convolutional layer 
% , as shown in equation \ref{eqn:mask_generates_2_masks}, 
to generate the real and imaginary components of the T-F mask, $\textbf{M} \in \mathbb{C}^{F \times T}$, with the same dimensions as a single-channel input. 
The mask is then applied to a reference channel, resulting in an enhanced single-channel speech representation, which can be represented as:
% Mask Predicting Multiplication
\begin{equation} \label{eqn:eq_mask_prediction_multiplication}
    \hat{\textbf{S}}(f,t) = \textbf{X}_{ref}(f,t) * \textbf{M}(f,t)\
\end{equation}

where $\hat{\textbf{S}} \in \mathbb{C} ^ {F \times T}$ is the STFT of the enhanced speech, $\textbf{X}_{ref} \in \mathbb{C} ^ {F \times T}$ is the STFT of the reference input signal, and $*$ is the complex multiplication operation. Finally, the enhanced representation is converted back into the time-domain speech signal using the inverse-STFT. 

\subsection{Bottleneck Network}
We experiment with integrating two types of Graph Neural Networks (GNNs) into the bottleneck of the U-Net:  Graph Convolutional Networks (GCNs) and Graph Attention Netoworks (GATs).  
We construct a graph, $\mathcal{G} = (\mathcal{V}, \mathcal{E})$, with $M$ nodes $v_i \in \mathcal{V}$, edges $(v_i, v_j) \in \mathcal{E}$, adjacency matrix $\mathcal{A} \in \mathbb{R}^{M \times M}$, and node degree matrix $\mathcal{D} \in \mathbb{R}^{M \times M}$, which is a diagonal matrix where $\mathcal{D}_{ii} = \sum_j A_{ij}$. 
The input node features to the GNNs are vectorized embedded representations of each channel $vec(\boldsymbol\zeta_i) \in \mathcal{R}^{d\Tilde{F} \Tilde{T}}$. 

The GNN's outputs are then reshaped to match encoder's output dimensions and passed as inputs to the decoder.
STFT.

\begin{table*}[]
\centering
\scalebox{0.8}{
\begin{tabular}{|c|c|ccccc|ccccc|}
\hline
\multirow{2}{*}{Dataset}   & \multirow{2}{*}{Method} & \multicolumn{5}{c|}{PESQ}                                                                                                                                              & \multicolumn{5}{c|}{STOI}                                                                                                                                              \\ \cline{3-12} 
                           &                         & \multicolumn{1}{c|}{BUS}            & \multicolumn{1}{c|}{CAF}            & \multicolumn{1}{c|}{PED}            & \multicolumn{1}{c|}{STR}            & AVG            & \multicolumn{1}{c|}{BUS}            & \multicolumn{1}{c|}{CAF}            & \multicolumn{1}{c|}{PED}            & \multicolumn{1}{c|}{STR}            & AVG            \\ \hline
\multirow{3}{*}{Simulated} & Noisy (channel 5)            & \multicolumn{1}{c|}{1.301}          & \multicolumn{1}{c|}{1.23}           & \multicolumn{1}{c|}{1.248}          & \multicolumn{1}{c|}{1.264}          & 1.261          & \multicolumn{1}{c|}{0.883}          & \multicolumn{1}{c|}{0.855}          & \multicolumn{1}{c|}{0.872}          & \multicolumn{1}{c|}{0.863}          & 0.868          \\
                           & U-Net Conv              & \multicolumn{1}{c|}{1.684}          & \multicolumn{1}{c|}{1.512}          & \multicolumn{1}{c|}{1.526}          & \multicolumn{1}{c|}{1.584}          & 1.576          & \multicolumn{1}{c|}{0.907}          & \multicolumn{1}{c|}{0.888}          & \multicolumn{1}{c|}{0.896}          & \multicolumn{1}{c|}{0.889}          & 0.895          \\
                           & RelUNet Conv            & \multicolumn{1}{c|}{\textbf{1.939}} & \multicolumn{1}{c|}{\textbf{1.81}}  & \multicolumn{1}{c|}{\textbf{1.762}} & \multicolumn{1}{c|}{\textbf{1.723}} & \textbf{1.809} & \multicolumn{1}{c|}{\textbf{0.942}} & \multicolumn{1}{c|}{\textbf{0.936}} & \multicolumn{1}{c|}{\textbf{0.931}} & \multicolumn{1}{c|}{\textbf{0.923}} & \textbf{0.933} \\ \hline
\multirow{3}{*}{Real}      & Noisy (channel 5)            & \multicolumn{1}{c|}{1.191}          & \multicolumn{1}{c|}{1.435}          & \multicolumn{1}{c|}{\textbf{1.473}} & \multicolumn{1}{c|}{1.31}           & 1.352          & \multicolumn{1}{c|}{0.383}          & \multicolumn{1}{c|}{0.517}          & \multicolumn{1}{c|}{0.462}          & \multicolumn{1}{c|}{0.382}          & 0.436          \\ 
                           & U-Net Conv              & \multicolumn{1}{c|}{1.188}          & \multicolumn{1}{c|}{1.401}          & \multicolumn{1}{c|}{1.406}          & \multicolumn{1}{c|}{1.293}          & 1.322          & \multicolumn{1}{c|}{0.369}          & \multicolumn{1}{c|}{0.541}          & \multicolumn{1}{c|}{0.479}          & \multicolumn{1}{c|}{0.392}          & 0.445          \\
                           & RelUNet Conv            & \multicolumn{1}{c|}{\textbf{1.256}} & \multicolumn{1}{c|}{\textbf{1.509}} & \multicolumn{1}{c|}{1.467}          & \multicolumn{1}{c|}{\textbf{1.383}} & \textbf{1.404} & \multicolumn{1}{c|}{\textbf{0.411}} & \multicolumn{1}{c|}{\textbf{0.563}} & \multicolumn{1}{c|}{\textbf{0.501}} & \multicolumn{1}{c|}{\textbf{0.413}} & \textbf{0.472} \\ \hline
\end{tabular}
}
\caption{Evaluation results of different noise conditions on the simulated and real test sets of the CHiME-3 dataset.}
\label{tab:per_noise_combined}
\end{table*}

\begin{table}[]
    \centering\scalebox{0.8}{
    \begin{tabular}{|c|c|c|c|c|c|}
    \hline
        \multirow{2}{*}{Model} & \multirow{2}{*}{Bottleneck} & \multicolumn{2}{c|}{Multi-channel} & \multicolumn{2}{c|}{Single-channel} \\ \cline{3-6} 
         &  & PESQ & STOI & PESQ & STOI \\ \hline
        Noisy (channel 5) & - & 1.26 & 0.87 & 1.26 & 0.87 \\ \hline
        % UNet & - & 1.39 & 0.83 & 1.55 & 0.88 \\ \hline
        UNet & - & 1.58 & 0.89 & 1.50 & 0.87 \\ 
        RelUNet & - & \textbf{1.81} & \textbf{0.93} & 1.47 & 0.87 \\ \hline
        UNet & GAT & 1.56 & 0.88 & 1.49 & 0.87 \\ 
        RelUNet & GAT & \textbf{1.72} & \textbf{0.92} & 1.56 & 0.87 \\ \hline
        UNet & GCN & 1.50 & 0.87 & 1.45 & 0.87 \\ 
        RelUNet & GCN & \textbf{1.73} & \textbf{0.91} & 1.56 & 0.87 \\ \hline
    \end{tabular}
    }
    \caption{Performance comparison of different models on CHiME-3 simulated test set for both multichannel and single-channel scenarios.}
    \label{tab:CHiME_simulated_combined}
\end{table}

\section{Experimental Setup}

\subsection{Datasets}

%To evaluate the performance of our proposed method, we used the CHiME-3. 
The CHiME-3 dataset \cite{barker2015chime3} provides real and simulated recordings using a 6-channel microphone array in four noisy environments: cafés (CAF), buses (BUS), pedestrian areas (PED), and streets (STR) sampled at 16 kHz. In this work, we used only the simulated data for training. The dataset includes 7,138, 1,640, and 1,320 utterances, corresponding to training, development, and test data, respectively. The simulated sets were used for training, development, and testing. Among the real sets, we only used the test set for evaluation.

% \subsection{Preprocessing}

The multichannel time-domain model input was peak-normalized using the peak value found among all channels. The single channel target was also peak normalized. Each sound file was split into 1.2-second segments, providing the data samples used to train, validate, and test the models. To compute the STFTs, we use a Fast Fourier Transform (FFT) length of 1024, a hop length of 151, and a Hanning window length of 1,024, where the last frequency bin is discarded.

\subsection{Network Architecture and Training}

We experimented with several model configurations, using a naming system where U-Net and RelUNet (ours) denote the backbone architecture, GCN and GAT refer to the integrated GNN in the bottleneck, and Conv. refers to the mask generation network. 
The backbone architectures consist of an encoder with six downsampling layers and a decoder with six upsampling layers. Each layer in the encoder and decoder uses convolutional blocks with SeLU activation \cite{klambauer2017selfnormalizing} and batch normalization. The mask generation network consists of a convolutional layer followed by SeLU activation. 
For the variations involving GNNs, we inserted two layers between the encoder and decoder. For the GAT, we used $1$ attention head. 

Models were trained using the Adam optimizer with a learning rate of 0.0001.  Training was conducted for 100 epochs with a batch size of 32, where the validation loss was computed after every training step to choose the best performing model. 
The used loss function is defined as:

\begin{equation}
    \mathcal{L} = 2 \cdot \| \hat{s} - s \| + \| \hat{M} - M \|
\end{equation}

where \( s \) and \( M \) are the time-domain signal and magnitude spectrogram, respectively, and $\hat{(.)}$ denote predicted values.

\section{Results and Discussion}
%\subsection{Evaluation Metrics}

To evaluate our method, we used Perceptual Evaluation of Speech Quality (PESQ) and Short-Time Objective Intelligibility (STOI). These metrics were computed using torch metrics\cite{detlefsen2022torchmetrics}; results are shown in Table \ref{tab:CHiME_simulated_combined}. 
We also report the performance with single-channel input, where the single channel is stacked with itself for the RelUNet model.

The detailed results for different noise types using the U-Net and RelUNet architectures tested on the simulated and real test sets are summarized in Tables \ref{tab:per_noise_combined}. The tables highlight the best performance for each metric in bold.

\subsection{Discussion}

\begin{figure}[h]
\centering
\includegraphics[width=0.45\textwidth]{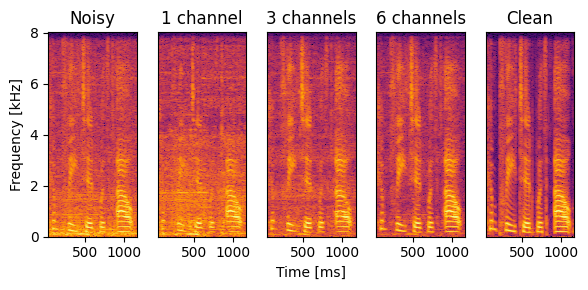}
\caption{Spectrograms illustration of the RelUNet Conv. using different number of channels.}
\label{fig:spectrograms_diff_channels}
\end{figure}

The multichannel results presented in Table \ref{tab:CHiME_simulated_combined} demonstrate that using a RelUNet backbone consistently enhances speech quality across various network architectures compared to the standard U-Net. Notably, while GNNs did not outperform other architectures, this can likely be attributed to the dataset's fixed speaker positioning, a result also reported in previous studies \cite{hao2022spatial}. Despite this, the GNNs still benefited from the RelUNet backbone, showing significant performance improvements, highlighting the robustness of RelUNet in boosting speech enhancement capabilities. The spectrograms in Figure \ref{fig:spectrograms_diff_channels} illustrates how the RelUNet leverages multichannel data. As the number of channels increases, the clarity of enhancement becomes more noticeable, particularly in regions where speech and noise are intertwined in both time and frequency. 

Although the RelUNet is primarily designed to enhance cross-channel information using a reference channel in a multichannel setup, it still demonstrates performance improvements in terms of PESQ in single-channel settings, as shown in Table \ref{tab:CHiME_simulated_combined}. 
This result aligns with traditional speech enhancement approaches that utilize autocorrelation methods for single-channel speech enhancement.

The performance of the models across different noise environments, as summarized in Table \ref{tab:per_noise_combined}, highlights the relative improvement offered by the RelUNet, compared to standard U-Net architecture used in previours neural speech enhancement research. In the real data, the U-Net model struggles to enhance speech and frequently degrades speech quality, showing poor generalization in this setting. The RelUNet, on the other hand, consistently improves over the U-Net and the noisy baseline. Note that in this case, both models are only trained on the simulated data, which is a common practice in speech enhancement research due to the difficulty of obtaining ground-truth clean signal from real noisy settings. The results demonstrate that the RelUNet has enhanced generalization ability to unseen noise conditions. Remarkably, the RelUNet achieves these significant improvements with only $\sim$0.07\% increase in number of parameters over the U-Net.

\section{Conclusions and Future Work}

This paper presented a novel approach to multichannel speech enhancement, called a RelUNet, leveraging a U-Net architecture combined with spatial feature integration via channel stacking . The proposed method significantly improves speech quality metrics, as demonstrated by our experimental results on the CHiME-3 dataset, with a negligible increase in number of parameters. The approach demonstrates consistent improvements across model architectures and even in single-channel settings. 
While our experiments with GNNs did not show enhanced improvements over a standard RelUNet, Future work will explore their applicability in  
data with variable microphone geometries.

\bibliographystyle{IEEEtran}  
\bibliography{refs}

\end{document}